\documentclass[11pt, oneside]{article}   	% use "amsart" instead of "article" for AMSLaTeX format
\usepackage{geometry}                		% See geometry.pdf to learn the layout options. There are lots.
\newcommand{\Lagr}{\mathcal{L}}
\usepackage{graphicx}				% Use pdf, png, jpg, or eps§ with pdflatex; use eps in DVI mode
\usepackage{amssymb}

\title{{\bf \small MiniBooNE Technical Note 307\\}
$\chi^2$ Fitting When Overall Normalization is a Fit Parameter}
\author{Byron Roe\\
Department of Physics\\
University of Michigan\\
Ann Arbor, MI 48109}
\begin{document}
%\end{document}
\maketitle

\section{Introduction}
The problem of fitting an event distribution when the total expected number of events is not fixed,
keeps appearing in experimental studies. 
Peelle's Pertinnent Puzzle (PPP) notes that in a $\chi^2$ fit, if overall normalization is one of the
parameters parameters to be fit, the fitted curve may be seriously low with respect to the
data points, sometimes below all of them.  This puzzle was the subject of a NIM article
by G. D'Agostini (NIMA 346 (1994) 306).  This problem and the solution for it are
well known within the statistics community, but, apparently, not well known among some of the physics 
community.  The purpose of this note is didactic, to explain the cause of the problem and the
easy and elegant solution.

The solution is to use maximum likelihood (ML) instead of $\chi^2$.  The essential
difference between the two approaches is that ML uses the normalization
of each term in the $\chi^2$ assuming it is a normal distribution, $1/\sqrt{2\pi\sigma^2}$.
In addition, the normalization is applied to the theoretical expectation not to the data.
In the present note we illustrate what goes wrong and how maximum likelihood fixes the problem in a very simple toy example which illustrates 
the problem clearly and is the appropriate physics model for
event histograms. We then note how a simple modification to the $\chi^2$ method gives
a result identical to the ML method.  I will also discuss the models in
G. d'Agostini's article (p. 309) and add one more. 

\section{Toy Model--$\chi^2$}
Consider a simple data set with only two bins.  Theory predicts that the expected value of
$N$, the number of events in the bin should be the same for each bin, and that the bins
are uncorrelated.  Let $x_1$ and $x_2$ be the number of events experimentally found in
the two bins. The variance ($\sigma^2$) is $N$ for each bin, ($\sigma=\sqrt{N}$).
\begin{equation}
\chi^2 = \frac{(N-x_1)^2}{\sigma^2} + \frac{(N-x_2)^2}{\sigma^2}.
\end{equation}
We want to find the minimum, $\frac{\partial\chi^2}{\partial N}= 0$. 
Call term 1, the derivative with
respect to the numerators of the $\chi^2$.
\begin{equation}
\rm{Term\ 1} = 2\frac{(N-x_1 +N-x_2)}{N} = 2\big(1-\frac{x_1}{N}\big) +2\big(1-\frac{x_2}{N}\big). 
\end{equation}
If we ignore the derivative of the denominator, then
Term 1 = 0, is solved by $N = \frac{x_1+x_2}{2}$.  Call this the naive solution.

Call Term 2 the derivative with respect to the denominator of the $\chi^2$
\begin{equation}
\rm {Term\ 2} = -\frac{(N-x_1)^2 +(N-x_2)^2}{N^2}.
\end{equation}
Term 2 is negative and O(1/N).  The only way that Term 1 + Term 2 = 0 is for
Term 1 to be positive.  This means that the $\chi^2$ solution must have $N$ greater than the
naive value.  Although Term 1 is O(1), $x_1/N$ and $x_2/N$ are O(1/N).  N is pulled up as the fit wants to make the fractional errors larger.  (Had the normalization been put into the data
not the theoretical value, the  fitted curve would have been low.)

\section{Toy Model--Maximum Likelihood}
The likelihood ($\Lagr$) is the probability density function for the two bins assuming each bin
has a normal distribution. (This requires $N$ is not too small).
\begin{equation}
\Lagr = \frac{1}{\sqrt{2\pi\sigma^2}} \frac{1}{\sqrt{2\pi\sigma^2}} 
e^{-(N-x_1)^2/(2\sigma^2)}e^{-(N-x_2)^2/(2\sigma^2)}.
\end{equation}
For $\sigma^2 = N$, the log of the likelihood is:
\begin{equation}
\ln\Lagr = -\ln(2\pi)-\ln N -\chi^2/2.
\end{equation}
Let Term 3 be the derivative of the normalization.  
\begin{equation}
\rm {Term\  3 }= -\frac{1}{N}.
\end{equation}
The derivative of the $\ln\Lagr$ is Term 3 $-$ (Term 1)/2 $-$ (Term 2)/2.
\begin{equation}
{\rm Term\ 3 - (Term\ 2)/2} = -\frac{1}{N} +  \frac{(N-x_1)^2 +(N-x_2)^2}{2N^2}
=\frac{ -2N +(N-x_1)^2 +(N-x_2)^2}{2N^2}.
\end{equation}
Since the expectation value  $E(N-x_1)^2 = E(N-x_2)^2 = N$ , the expectation value
of Term 3 - (Term 2)/2 =0.  For fitted values a modification is needed.  Assume that there
is only one overall normalization factor and assume now that there are $n_b$ bins. The expectation
value for a $\chi^2$ with $n_b$ bins and $n_f$ fitted parameters is $n_b -n_f$.  This
occurs because, after fitting, the multidimensional normal distribution loses $n_f$ variables.
This means, for $n_b=2,\ n_f=1$, the value of Term 2 is $2\times 1/2 = 1$.  The same loss
in dimensions requires term 3, the normalization term of the multidimensional distribution  to be multiplied by $(n_b - n_f)/n_b$ to match the change in $\chi^2$ since the fit has integrated
over those variables.
The change in expectation value occurs automatically  in the fit, but the modification to Term 3
must be put in by hand. 
 
There is an easy general way to handle this problem.  The problem arises
because the error matrix is a function of normalization.  When the simple $\chi^2$ method is
applied, the derivative of the $\chi^2$ is in error because the change in the normalization  of
the particle density function is not taken into account.  Including this term in the ML
approach eliminates the problem.  This leads to
a simple approach using a modified $\chi^2$ analysis.
Consider $n_b$ bins and $g$ fitting parameters $p_j$. 
Let $n_i(p_1,p_2,\cdots ,p_g)$
be the expected number of events in bin $i$.  The distribution of experimental events in each bin is taken as approximately normal.  The total number of events in the histogram is not fixed.
Choose the set $n_i$ as the basis.  The
error matrix is diagonal in this basis.  Ignoring
the $2\pi$ constants:
\begin{equation}
\ln\Lagr = \sum_{i=1}^{n_b} -\frac{\ln n_i}{2} -\frac{(x_i-n_i)^2}{2n_i}.
\end{equation}
\begin{equation}
\frac{d\ln\Lagr}{d n_i} = \frac{x_i-n_i}{n_i} +\frac{1}{2n_i}\big[\big( 
\frac{(x_i-n_i)^2}{n_i}\big)-1\big].
\end{equation}
The expectation value for the term in square brackets is zero.  Recall that the expectation
refers to the average value over a number of repetitions of the experiment.  It is $x_i$ that
changes with each experiment not the theoretical expectation, $n_i$.  The expectation value
of the term in square brackets will remain zero even if it
is multiplied by a complicated function of the $p_j$ fitting parameters.
Ignoring this term leads to:
\begin{equation}
\frac{\partial \ln\Lagr}{\partial p_j} = \sum_{i=1}^{n_b}  \big(\frac{x_i-n_i}{n_i}\big)\frac{\partial n_i}{\partial p_j}.
\end{equation}
By expressing the $n_i$  as the appropriate functions of the $p_j$, the error matrix
can be written in terms of the $p_j$.
However, the derivative of the inverse error matrix does not appear in the transform of Equation 10.
This result means that one can use a modified $\chi^2$ approach.  Use the usual $\chi^2$,
but, when derivatives are taken to find the $\chi^2$ minimum, omit the derivatives of the
inverse error matrix.  The result is identical to the result from ML.  The modified $\chi^2$
method should be generally used in place of the regular $\chi^2$ method.

In practice, since the differences are not precisely the expectation
values for a given experiment, there is a small residual higher order effect,
which causes no bias on the average.

%The procedure for
%finding the normalization of a multidimensional normal distribution with correlations is 
%straightforward and can be found in many statistics texts.
%In my book, ``Probability and Statistics in Experimental Physics, 2nd Edition" it is in 
%Chapter 10, pages
%99-103.

\section{Review of  G. D'Agostini's models}

The problem he discusses is a bit different than that treated in the toy model.  He
imagines that we have two measurements of the same physical quantity, but that
there is a possible scale error $f$ and a best value $k$ of two measurements, $x_1$ and $x_2$ to 
be fit.  
The models presented by D'Agostini can be written in the form:
\begin{equation}
\chi^2_n= \frac{(fx_1-k)^2}{f^n\sigma_1^2 }+ \frac{(fx_2-k)^2}{f^n\sigma_2^2} +\frac{(f-1)^2}{\sigma_f^2}
= \frac{(x_1 -k/f)^2}{f^{n-2}\sigma_1^2} + \frac{(x_2-k/f)^2}{f^{n-2}\sigma_2^2} + \frac{(f-1)^2}{\sigma_f^2}.
\end{equation}
He treats the cases n=2 (Model A) and n=0 (Model B).  We will also discuss the
case $n=-1$.  D'Agostini finds that $n=2$ does not exhibit PPP, but $n=0$ does
exhibit it.

There are two errors in the method of D'Agostini, which we have already mentioned in the
previous section.
\begin{itemize}
\item The use of the $\chi^2$ distribution incorrectly ignores the changes of normalization
of the multidimensional density distribution as the normalization parameter is changed.
\item The normalization parameter $N$ should be included in the theoretically expected
value, not in the data value.  The experimentally observed number of events is what it is.
D'Agostini's $f = 1/N$.  This has two effects.  The first effect is that  the 
normalization dependence of the error matrix is changed.  The second effect is that the
average of $N$ is not the same as the average of $1/N$. 
\end{itemize}

First consider the ML solution.  Using $N$ as normalization,
\begin{equation}
\chi^2 = \frac{(x_1-Nk)^2}{N^{2-n}\sigma_1^2} + \frac{(x_2-Nk)^2}{N^{2-n}\sigma_2^2} 
+\frac{(N-1)^2}{\sigma_N^2}.
\end{equation}
It is assumed here that $\sigma_N^2$ is a fixed number, rather than having $\sigma_f^2$ fixed.
Let 
\begin{equation}
\chi^{2*} =\chi^2 -\frac{(N-1)^2}{\sigma_N^2}.
\end{equation}
The derivative of the numerator of $\chi^2$ with respect to $N$ is:
\begin{equation}
\frac{2(Nk-x_1)}{N^{2-n}\sigma_1^2} + \frac{2(Nk-x_2)}{N^{2-n}\sigma_2^2} 
+\frac{2(N-1)}{\sigma^2_N}.
\end{equation}
The derivative of the denominator is:   
\begin{equation}
\frac{n-2}{N}\chi^{2*}.
\end{equation}    
For ML the $N$ dependent part of the normalization term is $(1/\sqrt{N^{2-n}})^2$.
The log of this term is $-(2-n)\ln N$ and the derivative of the log with respect to $N$ is $(n-2)/N$.
For ML then:
\begin{equation}
\frac{\partial {\rm ML}}{\partial N} = \frac{n-2}{N} - \frac{1}{2}\big( \frac{2(Nk-x_1)}{N^{2-n}\sigma_1^2} + \frac{2(Nk-x_2)}{N^{2-n}\sigma_2^2} +\frac{2(N-1)}{\sigma^2_N}+   
\frac{(n-2)}{N}\chi^{2*}\big).
\end{equation}
%-\frac{2}{N}\frac{(n-1)^2}{N^2\sigma_f^2}\big).$$

Here, the expectation value of the $\chi^{2*}$ term  is 1 after fitting and the
normalization term is reduced to $(n-2)/(2N)$ to account for the loss of a degree of
freedom.  For any $n$, the ML normalization term 
cancels the expectation value of the denominator derivative.

Next look at this using D'Agostini's calclulation.
For any $n$ value, the derivative with respect to $k$ is:
\begin{equation}
\frac{\partial \chi^2_n}{\partial k} = \frac{2}{f^{n-1}}[\frac{(k/f-x_1)}{\sigma_1^2}
+ \frac{(k/f-x_2)}{\sigma_2^2}]=0.
\end{equation}
Hence,
\begin{equation}
 k =f(\frac{x_1}{\sigma_1^2}+\frac{x_2}{\sigma_2^2})/(\frac{1}{\sigma_1^2}+\frac{1}{\sigma_2^2}),
 \end{equation}
which is the expected result from combining two measurements of the same quantity, except
for the factor $f$.   
Define the result for $f=1$ to be $\overline{x} $.
\begin{equation}
\overline{x} = (\frac{x_1}{\sigma_1^2}+\frac{x_2}{\sigma_2^2})/(\frac{1}{\sigma_1^2}+\frac{1}{\sigma_2^2}),
\end{equation}

Note that for $\frac{\partial \chi^2}{\partial f}$, the derivative of the numerators of the first two
terms together (using $\frac{(fx_1-k)^2}{f^n\sigma_1^2 }+ \frac{(fx_2-k)^2}{f^n\sigma_2^2}) $
has been determined to be zero from the $\frac{\partial \chi^2}{\partial k}$ derivative.

\subsection{$n=2$, Model A}
 Using the result from the derivative with respect to $k$, it is seen that
for the derivative with respect to $f$, (using the 2nd expression  in Equation 11 with 
$f^{n-2}=1$ in the
denominator), the derivatives of the first two terms
add to be zero from the result of the derivative with respect to $k$ seen in Equation 17, and then $f$ is forced to be 1.  D'Agostini finds that this does not have
a PPP problem as expected since the variance is independent of $f$.

%\subsection{$n=1$}
%The first two terms of the numerators of
%the $\chi^2$ derivative are zero for the derivative with respect to $f$.  Then:
%$$\frac{\partial \chi^2_{n=1} }{\partial f} = -\frac{1}{f}[\frac{(fx_1-k)^2}{f\sigma_1^2}
%+ \frac{fx_2 -k)^2}{f\sigma_2^2} ]+   \frac{2(f-1)}{\sigma^2_f}. $$
%Note that the expectation value of the bracketed terms is two so
%$$\frac{\partial \chi^2_{n=1} }{\partial f} \approx -\frac{2}{f} +\frac{2(f-1)}{\sigma^2_f}. $$

\subsection{$n=0$, Model B}
\begin{equation}
\chi^2_B = \frac{(fx_1-k)^2}{\sigma_1^2 }+ \frac{(fx_2-k^2)}{\sigma_2^2} +\frac{(f-1)^2}{\sigma_f^2}
= \frac{(x_1 -k/f)^2}{\sigma_1^2} + \frac{(x_2-k/f)^2}{\sigma_2^2} + \frac{(f-1)^2}{\sigma_f^2}.
\end{equation}
\begin{equation}
\frac{\partial \chi^2_B}{\partial k} = 2[\frac{(k -fx_1)}{\sigma_1^2}
+ 2[\frac{(k-fx_2)}{\sigma_2^2}].
\end{equation}
Here, $f$ will not be one.  Using the result from the partial derivative with respect to $k$,
$\chi^2_B$ can be written:
\begin{equation}
 \frac{\partial \chi^2_{B}}{\partial f} = 2f^2[\frac{(x_1-\overline{x})^2}{\sigma_1^2} +\frac{(x_2-\overline{x})^2}{\sigma_2^2}]
+2\frac{(f-1)}{\sigma^2_f}.
\end{equation}
\begin{equation}
\frac{1}{f} = \sigma_f^2[\frac{1}{\sigma_f^2} +\frac{(x_1-\overline{x})^2}{\sigma_1^2}
+\frac{(x_2-\overline{x})^2}{\sigma_2^2}].
\end{equation}
%Let $\hat{x} =f\overline{x}$:
%Then 
\begin{equation}
%r=\frac{\hat{x}}{\overline{x}} 
f= 1/\big[1+\sigma_f^2\big(\frac{(x_1-\overline{x})^2}{\sigma_1^2} +\frac{(x_2-\overline{x})^2}{\sigma_2^2}  \big)\big].
\end{equation}
\begin{equation}
x_1-\overline{x} = x_1 - (\frac{x_1}{\sigma_1^2} +\frac{x_2}{\sigma_2^2} )
/(\frac{1}{\sigma_1^2} + \frac{1}{\sigma_2^2}) =\frac{x_1-x_2}{\sigma_2^2(1/\sigma_1^2+1/\sigma_2^2)}.
\end{equation}
Similarly,
$$x_2-\overline{x} = \frac{x_2-x_1}{\sigma_1^2(1/\sigma_1^2 +1/\sigma_2^2)}.$$
To find $f$, consider:
\begin{equation}
\frac{(x_1-\overline{x})^2}{\sigma_1^2} + \frac{(x_2-\overline{x})^2}{\sigma_2^2} =
\frac{(x_1-x_2)^2}{\sigma_1^2\sigma_2^4(1/\sigma_1^2+\sigma_2^2)^2}
+ \frac{(x_1-x_2)^2}{\sigma_1^4\sigma_2^2(1/\sigma_1^2+\sigma_2^2)^2}
=\frac{(x_1-x_2)^2}{\sigma_1^2+\sigma_2^2}.
\end{equation}
\begin{equation}
f= \frac{1}{1+\sigma_f^2(x_1-x_2)^2/(\sigma_1^2+\sigma_2^2)}.
\end{equation}
$f$ is always less than one. This  is the result obtained by D'Agostini.

\subsection{$n=-1$, the Toy Model}
Use the notation of D'Agostini.  Again the first two terms of $\frac{\partial \chi^2_{n=-1}}{\partial f}$
are zero.
\begin{equation}
\frac{\partial \chi^2_{n=-1}}{\partial f} = \frac{1}{f}[\frac{(fx_1-k)^2}{f^{-1}\sigma_1^2}
+ \frac{(fx_2-k)^2}{f^{-1}\sigma_2^2}]+\frac{2(f-1)}{\sigma_f^2}.
\end{equation}
The expectation value of the first two terms is $\frac{2}{f}$.
\begin{equation}
\frac{\partial \chi^2_{n=-1}}{\partial f} \approx \frac{2}{f}+\frac{2(f-1)}{\sigma_f^2}.
\end{equation}
This will be far from $f=1$, unless $\sigma_f<<1$.
However, the ML term is $\frac{1}{f}$.
\begin{equation}
 \frac{\partial \ln\Lagr_{n=-1}}{\partial f} 
= \frac{1}{f}-\frac{\chi^2_{n=-1}}{2}\approx \frac{1}{f}-\frac{1}{f} 
-\frac{(f-1)}{\sigma_f^2}.
\end{equation}
For the ML method, $f = 1$.

\section{Summary}
The PPP problem arises because the $\chi^2$ method incorrectly ignores the normalizations
of the multidimensional probability density functions when the total expected number of events is not fixed.
For an event histogram the maximum likelihood method is correct if:
\begin{itemize}
\item{} Errors are taken as the square root of the theory model; they are not to be
taken as the square root of the number of events in the bin.
\item{} The normalization factor is included with the theory model.  
\item{} The subtraction for noise is included with the theory model.The data is
the number of events obtained experimentally.  All corrections belong to the theory model.
\end{itemize}
This ML result is completely equivalent to a modified $\chi^2$ approach.  Use the usual $\chi^2$,
but, when derivatives are taken to find the $\chi^2$ minimum, omit the derivatives of the
inverse error matrix. 
\end{document}